# Quark-parton Fusion and the Scaling of Hadron Production at the LHC

Alexey B. Kurepin*, Evgeny V. Karpechev,

Institute for Nuclear Research, Moscow, Russia

Email: *kurepin@inr.ru

## Abstract

Hadron production, including pions, kaons and antiprotons, in proton-proton collisions at an energy of √s = 7 TeV at the ALICE facility has been studied. For the data analysis, a generalised parton model was used, which takes into account the interaction of two partons in colliding beams and the production of new particles. Considering the effects of parton fusion and compound state formation results in a scaling dependence of the hadron production cross sections of different species.

## Keywords

Ultrarelativistic proton-proton collisions, Hadronic production, Scaling, Parton fusion

## 1. Introduction

The experimental and theoretical study of hadron production is one of the main areas of research into the state of nuclear matter in proton-proton and proton-ultrahigh-energy-nucleus collisions at the LHC. Several studies have been conducted at the ALICE facility to determine the production cross sections of pions, kaons, protons and antiprotons in proton-proton interactions at energies of 0.9 TeV [1], 7 TeV [2,3], and 13 TeV [4]. Hadron production was also measured in lead nucleus collisions at energies of 2.76 TeV [5].

Particle identification was performed using three methods: the determination of specific energy losses for ionisation, as measured by the ITS and TPC; time-of-flight, as measured by the TOF detector; and Cherenkov emission, as measured by the High-Momentum Particle Identification Detector (HMPID). Combining these methods provides precise measurements of $p_{\mathrm{T}}$ spectra over a wide range of momenta: 0.1–3 GeV/c for pions, 0.2–6 GeV/c for kaons, and 0.3–6 GeV/c for protons and antiprotons. Comparing the ALICE results with similar measurements performed by the PHENIX collaboration at RHIC shows that the pT-integrated yields increase with collision energy for all measured particle species. A slight increase in the mean pT values with increasing energy is also observed. This may be due to the increasing contribution of hard processes at these energies [6].

Hadron formation mechanisms are typically studied by comparing measured transverse momentum distributions for pions, kaons, protons and antiprotons with calculations using Monte Carlo event generators based on quantum chromodynamics (QCD). PYTHIA [7,8], EPOSLHC [9] and PHOJET [10] generators are the most commonly used to describe proton-proton collisions at high energies. As it was shown, the PYTHIA and EPOS generators can describe the pion transverse momentum distributions over the entire $p_{\mathrm{T}}$ range with an accuracy of about 15%. However, the PHOJET generator does not provide a satisfactory description of the measured shape of the spectrum for any of the particle species. Deviations from the data peak at $p_{\mathrm{T}} \simeq 1.2$ GeV/c, being more pronounced for kaons and protons than for pions. All Monte Carlo generators underestimate the kaon yield by 20–30% for $p_{\mathrm{T}} > 600$ MeV/c and overestimate it by up to 30% for $p_{\mathrm{T}} < 400$ MeV/c. EPOS only accurately describes the proton yield at low transverse momenta ($p_{\mathrm{T}} < 1$ GeV/c); at higher $p_{\mathrm{T}}$, the generator tends to overestimate the data by up to 30%. PYTHIA does not accurately describe the shape of the proton spectrum across the entire $p_{\mathrm{T}}$ range. A reasonable description of the proton yield is obtained in the range $1 < p_{\mathrm{T}} < 2$ GeV/c, but the data at lower and higher $p_{\mathrm{T}}$ are overestimated by up to 40%. None of the generators can simultaneously describe the measured yields of pions, kaons, protons and antiprotons (Fig. 1). Comparing the data with Monte Carlo calculations shows that tuning the generators based on just a few global observables, such as the integrated multiplicity of charged hadrons, only provides a partial description of the data.

While the shapes of the spectra are reasonably well reproduced by all models (except PHOJET, which does not describe the spectrum shapes of all three hadron varieties), none can simultaneously describe the measured yields of pions, kaons and protons. These results can be used to improve our understanding of hadronic formation mechanisms in pp interactions at LHC energies, and may further constrain the parameters of the models.

The proposed z-scaling model [11] was developed in an attempt to describe inclusive particle production. In this model, a z value is introduced instead of the known Bjorken scaling parameters, and this value also takes into account dynamical variables. Using this representation, the production of pions in proton-nucleus interactions at energies ranging from 70 to 400 GeV could be described by a single universal curve. However, data on kaon and antiproton production do not fit this universal curve [12].

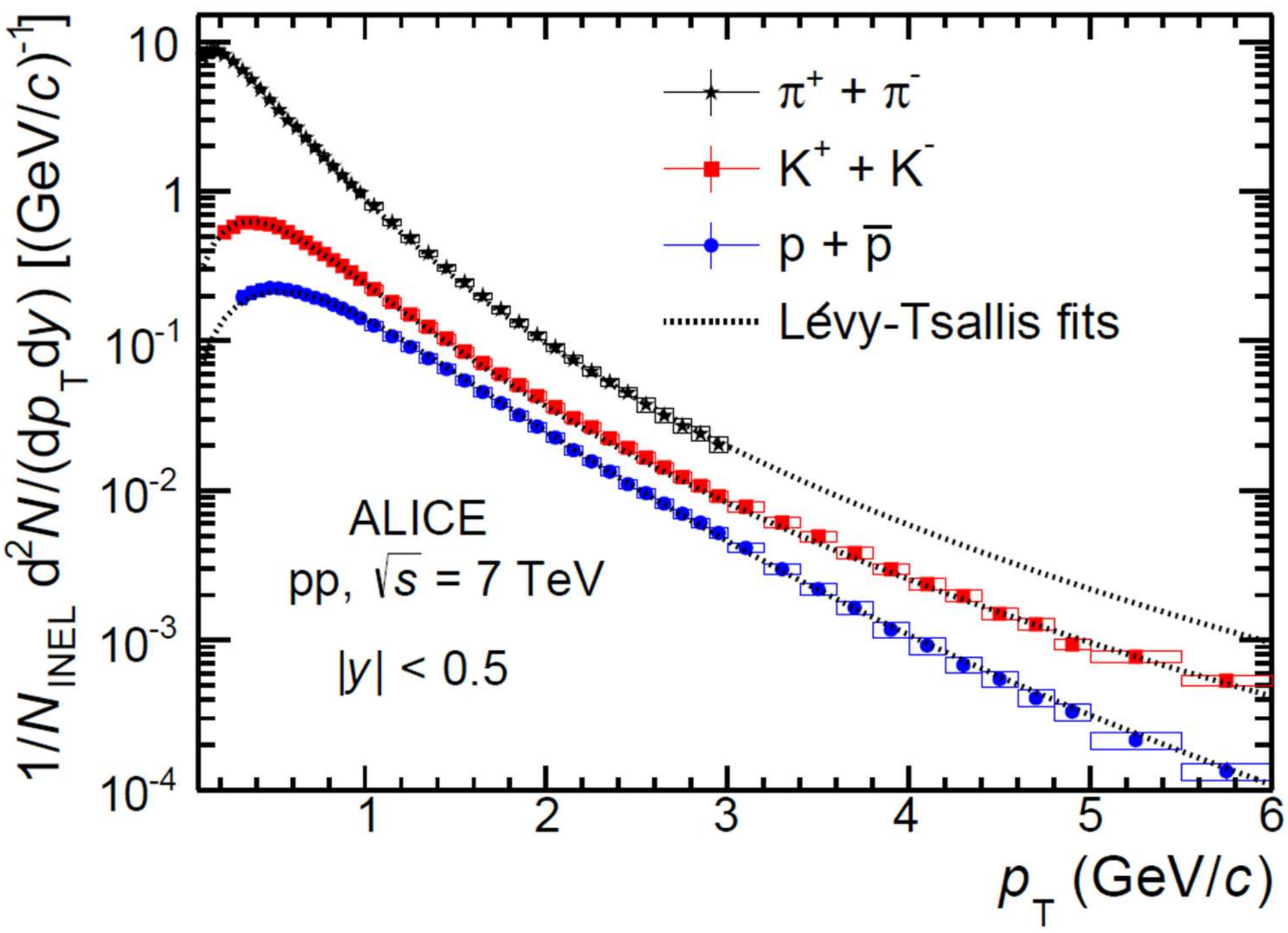

Fig. 1. Combined $p_T$ spectra of π, K and p (sum of particles and antiparticles), measured with ALICE at midrapidity (y < 0.5) in pp collisions at √s = 7 TeV [2]

Thus, there are no theoretical models for hadron production in proton-proton interactions at high energies that can explain the differences in yields of pions, kaons, protons and antiprotons, on the observed transverse momentum spectra (Fig. 1). Conversely, in the simpler process of deep inelastic scattering of electrons on protons, scaling with the Bjorken parameter (representing the parton momentum fraction in the nucleon) was found, providing justification for the quark model [13]

The process of hadron production is more complicated than deep inelastic scattering because it involves the production of new particles and multiple processes. Nevertheless, it is evident that the quark-parton interaction mechanism is also involved in hadron production during the collision of high-energy protons. In this case, the scaling result may be a universal dependence on the parton four-momentum, equal to the fraction of the four-momentum of the colliding protons. This work demonstrates that the production of pions, kaons, protons and antiprotons at the ALICE facility at a proton energy of √s = 7 TeV can be described by a universal function of the scaling parameter x, provided that parton fusion occurs as a result of the formation and subsequent decay of a compound object, with hadrons being emitted.

A new analysis of hadron production was made. The analyzed data represent proton–proton collisions at √s = 7 TeV, collected in 2010. Events were selected from tracks reconstructed in the TPC and ITS. Global tracks must intersect at least 70 TPC readout lines with an acceptable $\chi^2$ momentum fit in the TPC and have at least two clusters reconstructed in the ITS. Full details of the ALICE experiment and detectors can be found in [2].

## 2. The Parton Fusion Model and Scaling of Hadron Production at the LHC.

Quark fusion in the collision of protons and nuclei at high energies may be one of the mechanisms of hadron production. Since the energies of the hadrons detected at the LHC are much lower than those of the colliding partons, it can be assumed that the production mechanisms of different types of hadrons are the same, that the effective production cross sections are close and that the production rate is determined by the universal scaling parameter. A similar mechanism involving quark fusion and the formation of a compound state has recently been considered in the case of the doubly-charmed baryon $\Xi_{cc}^{++}$ which is formed by the fusion of two heavy baryons, the $\Lambda_c$ [14].

The merger of two partons results in the formation of a compound parton whose mass is equal to the sum of the masses of the two original partons, and whose longitudinal momentum is equal to the difference in momentum between the two colliding partons. The compound parton can exist in various excited states. Further decay of the compound parton leads to the formation of a hadron, which is observed in experiments, and a remnant, whose form is determined by the laws of conservation of baryon number and strangeness. For example, for the emission of an antiproton, the remnant is a proton; for the emission of a kaon, the remnant is a kaon with opposite charge.

The decay of a compound parton must lead to the emission of a hadron pair to compensate for the measured hadron's transverse momentum. Thus, the particle pair is emitted not only for antiprotons and kaons, but also for pions and protons. Therefore, due to momentum conservation the longitudinal momentum of the compound parton is equal to longitudinal momentum of produced particles:

$$( x_1 - x_2 ) P = 2 p_{3L}, \tag{1}$$

where $P$ is the proton's momentum in the collider and $p_{3L}$ is the hadron's longitudinal momentum.

Due to the conservation of 4-momentum, we have:

$$( x_1 q_1 + x_2 q_2 - q_3 )^2 = q_4^2 \tag{2}$$

where $q_1$ and $q_2$ are the 4-momenta of the colliding protons, $q_3$ is the 4-momentum of the produced hadron and $q_4$ is the 4-momentum of the parton remaining after the decay of the compound parton from the excited state with hadron emission and the additional particle required for the conservation of baryon number and strangeness.

Under the assumptions discussed above regarding parton fusion and decay of the compound parton, $q_4^2$ is equal to the square of the residue's mass.

$$M = x_1 m + x_2 m + m_i \quad (3)$$

where m is the proton mass and $m_i$ is the mass of the additional particle.

By squaring, replacing the total energy of the colliding protons with momentum and taking into account the equation (1) we obtain the equation for calculating the scaling parameter $x_2$ :

$$4P^2 x_2^2 + 4P(2P_3\cos\theta - E_3)x_2 + 4P_3\cos\theta(P_3\cos\theta - E_3) + m_3^2 - M^2 = 0 \quad (4)$$

where θ is the emission angle of the hadron, $E_3$ , $P_3$ and $m_3$ are the total energy, momentum and mass of the produced hadron, respectively.

Based on the kinematic data for each hadron type and particle yields obtained in the ALICE experiment at a proton energy of 7 TeV, we calculated the values of the parameter $x_2$ and the resulting distributions of particle yields with respect to this parameter, shown in Figure 2.

It turns out that all distributions are located on the universal curve, except in the threshold region. Thus, scaling is observed on the parameter $x_2$ (Fig. 2).

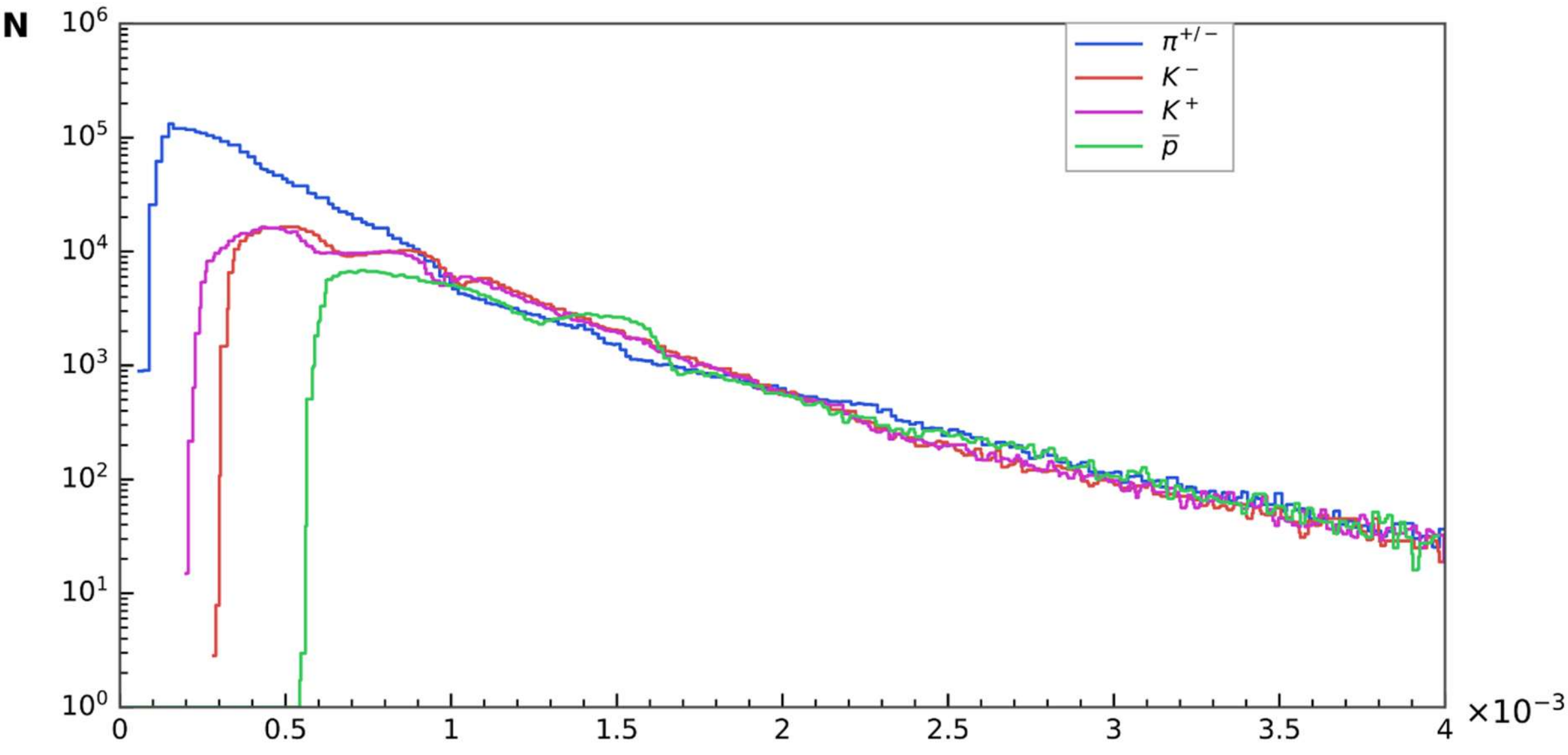

Fig. 2: Parton distribution of the $x_2$ parameter obtained by equation (4)

Statistical errors are of the order of the point size. Some deviations from a straight line are observed due to a problem with determining the total momentum. Excluding these deviations and the threshold region, the accuracy of the average scaling parameters for the hadrons under consideration $\chi^2/df$ is in the range of 0.2–0.5.

Due to the symmetry of the experiment with respect to the point at which the beams meet, the distributions of $x_1$ and $x_2$ must coincide. As many partons are involved in the collision, the process is inclusive and consists of many elementary hadron production reactions.

Since the experiment had limitations related to the registration of hadrons and spectrometry at several GeV the values of $x_1$ and $x_2$ are at the level $10^{-3}$.

A similar analysis can be performed for other baryons, such as Λ, Σ and Ξ, when complete kinematic data become available.

## 3. Conclusion

The example of hadron production at the ALICE facility shows that the quark-parton model applies not only to deep inelastic scattering, but also to the collision of two partons in a collider. In this interaction, a compound state is formed by parton fusion and subsequent hadron emission. Regardless of the hadron species, the production cross section is determined by scaling dependence.

## Declaration of Competing Interest

The authors declare that they have no competing financial interests or personal relationships that could have appeared to influence the work reported in this paper.